# Realization of room-temperature ferromagnetic semiconducting state in graphene monolayer

Yu Zhang[1, §], Xue-Lei Sui[2, §], Dong-Lin Ma[1], Ke-Ke Bai[1], Wen-Hui Duan[2], and Lin He[1,*]

[1]Center for Advanced Quantum Studies, Department of Physics, Beijing Normal University, Beijing, 100875, People's Republic of China

[2]State Key Laboratory of Low-Dimensional Quantum Physics and Collaborative Innovation Center of Quantum Matter, Department of Physics, Tsinghua University, Beijing 100084, People's Republic of China

Correspondence and requests for materials should be addressed to L.H. (e-mail: helin@bnu.edu.cn).

**Room-temperature ferromagnetic semiconductor is vital in nonvolatile digital circuits and it can provide an idea system where we can make use of both charge and spin of electrons. However, seeking room-temperature ferromagnetic semiconductors is still just an appealing idea that has never been realized in practice up to now. Here we demonstrate that graphene monolayer, hybridized with underlying Ni(111) substrate, is the room-temperature ferromagnetic semiconductor that has been continuously searched for decades. Our spin-polarized scanning tunnelling microscopy (STM) experiments, complemented by first-principles calculations, demonstrate explicitly that the interaction between graphene and the Ni substrate generates a large gap in graphene and simultaneously leads to a relatively shift between majority- and minority-spin bands. Consequently, the graphene sheet on the Ni substrate exhibits a spin-polarized gap with energy of several tens meV even at room-temperature. This result makes the science and applications of room-temperature ferromagnetic semiconductors achievable and raises hopes of graphene-based novel information technologies.**

Ferromagnetic semiconductors, taking great advantages of both charge and spin degree of freedom of electrons, open up possibilities for new semiconductor-based data processing and memory applications[1,2]. During the past decades, looking for high Curie-temperature ($T_C$) ferromagnetic semiconductors, especially for those with the ferromagnetism persisting above room-temperature, has attracted immense research interest. There are two general strategies to achieve this objective. A straightforward method is to directly synthesize gapped materials with high $T_C$ ferromagnetism. Stoichiometric EuO is one of the rare obtained intrinsic ferromagnetic semiconductors with the record high Curie temperature of ~ 69 K[3-5]. Although it was predicted to host many promising properties (such as the nearly 100% spin-polarized current), the relatively low $T_C$ and instability in air hinder its application[5]. Another available method is to introduce a high concentration of magnetic ions into nonmagnetic semiconductors[6]. According to the meanfield *p-d* Zener model[6], the $T_C$ of the diluted magnetic semiconductors (DMSs) depends tightly on the magnetic impurity concentration *x* and hole concentration *p* (carrier mobility) as $T_C \propto xp^{1/3}$. Several well-established DMSs[7-15], such as (In,Mn)As and (Ga,Mn)As, have been realized successfully and demonstrated experimentally to be robust ferromagnetic semiconductors. However, due to the low solubility of magnetic elements and the immature technology to highly concentrate the magnetic ions at present, the obtained highest $T_C$ of the ferromagnetic semiconductors is still lower than 200 K up to now. Therefore, after seeking for several decades with great efforts, the realization of room-temperature ferromagnetic semiconductors seems to become a science fiction that is beyond the grasp of today's technology.

Both the said two strategies try to induce room-temperature ferromagnetic semiconducting state in the whole ultimate material, as schematically shown in Fig. 1a. However, it was demonstrated to be almost impossible in practice. Here, we report a new recipe to achieve this goal by only driving a small part, i.e., only the surface, of an artificial system into the room-temperature ferromagnetic semiconducting state (Fig. 1b). Our work demonstrates that graphene monolayer on Ni(111) substrates is the room-temperature ferromagnetic semiconductor that scientists dream to realize it for decades. Pristine graphene is a diamagnetic semimetal[16]. Recently, many groups attempt to introduce magnetism in graphene[17-28]. For example, it has been demonstrated explicitly that graphene with atomic defects and hydrogen chemisorption defects could exhibit localized magnetic moments. However, it is almost impossible to realize macroscopic room-temperature spin-polarized semiconductor band structure in graphene by simply introducing adatoms or atomic vacancies. The recipe reported in this work could overcome this difficulty completely. Once synthesized graphene monolayer on Ni(111) substrate, the electronic structure of graphene π band is strongly perturbed by d electrons of Ni. The π-d interaction not only opens up a gap ($E_{\sigma\uparrow}$ and $E_{\sigma\downarrow}$ for the spin-up and spin-down electrons, respectively) of several hundreds meV in graphene, but also lifts the degeneracy of the majority- and minority-spin bands around K point of graphene[29], thus realizing the spin-polarized ferromagnetic semiconducting state with the gap of tens meV in graphene. Very recently, the local magnetic moments of atomic defects in graphene and the emergence of edge magnetism in individual zigzag graphene nanoribbon are studied successfully by measuring their electronic structures via scanning tunnelling microscopy (STM)[17-20,22-24]. Here, we use spin-polarized STM measurements, for the first time, to directly measure both the relatively energy shift between the majority- and minority-spin bands and the gap opening of the graphene monolayer on Ni(111) substrate. These results provide clearly signatures of the emergence of the room-temperature ferromagnetic semiconducting state in the graphene monolayer on Ni(111) substrate.

In this paper, the graphene monolayer was grown on a Ni(111) single crystal (5 mm ×5 mm width, 1 mm thickness) via a traditional low pressure chemical vapor deposition (LPCVD) method[30-32] (the growth process is schematically shown in Fig. S1). The thickness of the as-grown graphene is characterized by Raman spectra measurements. Figure 2a shows typical Raman spectra measured at different positions of the synthesized graphene transferred to a 300 nm $SiO_2$/Si substrate. Two peaks at 1580 cm$^{-1}$ and 2700 cm$^{-1}$ in the Raman spectra are the G-band and 2D-band respectively, which are the characteristic Raman peaks of graphene[30-33]. We can identify the layer number of graphene by comparing the intensities of the G-band and 2D-band. Figure 2b shows the values of $I_{2D}/I_G$ with the average value about 2, indicating that the obtained sample under the growth process is mainly graphene monolayer. Such a result is further confirmed by our STM measurements. Fig. 2c and 2d show a representative 100 nm × 100 nm STM topographic image of the graphene monolayer across typical monatomic Ni(111) steps. The widths of Ni(111) terraces vary from about 20 nm to about 80 nm. As shown in Fig. 2e, the atomic graphene lattice can be resolved on the terraces, exhibiting enormous intensity imbalance between the *A* and *B* sublattices. Such a result indicates the inversion symmetry breaking of graphene by the substrate. For graphene on Ni(111) surface, the carbon atoms of the *A* sublattice are mainly on top of the Ni atoms of the topmost atomic layer, while atoms of the *B* sublattice are mainly located above the hollow sites of the topmost atomic layer of Ni due to their minor lattice mismatch, as schematically shown in Fig. 2f. We will demonstrate subsequently that the strong chemical interaction between graphene and Ni substrate not only results in the enormous sublattice asymmetry, but also dramatically changes the electronic band structure of graphene.

Figure 3a shows several representative scanning tunneling spectroscopy (STS) spectra of the graphene monolayer on Ni(111) surface recorded at different temperatures. It is interesting to note that we clearly observe a finite gap, ~ 40 meV, in the graphene monolayer at low temperature. With increasing the temperature, the gap feature in the STS spectra becomes weak due to thermal broaden. However, we still can detect the gap-like feature even at room temperature (300 K). Our experiment demonstrates that the main features of the STS spectra measured at different terraces (and at different positions) are almost the same and irrespective of the width of the terraces (see Fig. S2). Moreover, similar spectra are obtained with several different nonmagnetic STM tips (chemical etching from a wire of Pt(80%) Ir(20%) alloys), which removes any possible artificial effects as the origin of the observed result. To further explore the origin of the observed gap for the graphene monolayer on Ni(111) surface, we carried out similar STS measurements on bare Ni(111) (see Fig. S3). The distinct spectra recorded in bare Ni(111) remove the electronic states of Ni(111) as the origin of the observed gap in Fig. 3a. Simultaneously, we also measured STS spectra on an insulating hexagonal boron nitride (hBN) monolayer on Cu foil for comparison. In the hBN/Cu system, we clearly observed the large band gap, ~ 5.9 eV, of the hBN monolayer (see Fig. S4). This indicates that the STM could predominantly probe the electronic states of the topmost monolayer underneath the STM tip, and the observed gap in Fig. 3a mainly reflects the electronic structure of the graphene monolayer on the Ni(111) surface[29,34]. The Ni(111) surface should play a vital role in the emergence of the gap in graphene because that graphene monolayer on other metallic substrate, such as Cu[35-38], usually exhibits V-shaped spectra and there is no measurable band gap (Fig. S5), as expected to be observed for the pristine graphene monolayer.

The observation of a finite gap in the graphene monolayer on the Ni(111) surface is quite reasonable since that the substrate generates enormous sublattice asymmetry in graphene (Fig. 2e), which is expected to open a gap in it[39-43]. However, with considering the strong chemical interaction between graphene and Ni and the observed large sublattice imbalance in graphene, the obtained gap, $E_{gap}$~40 meV, in the graphene monolayer (Fig. 3a) is unexpectedly small. The observed gap is only comparable to the expected gap for a graphene monolayer on hBN substrate[40,41], whereas the interaction between graphene and hBN is much weaker than that between graphene and Ni. To fully understand the electronic structure of the graphene monolayer on Ni (111) surface, we carried out first-principles calculations on this system. Fig. 3b shows a representative theoretical electronic band structure of the graphene monolayer on Ni(111) surface and the inset of Fig. 3b shows the calculated STM image of the graphene monolayer on Ni(111) surface. Obviously, the strong chemical interaction between graphene and Ni substrate results in the enormous sublattice asymmetry in graphene, as observed in our experiment. The enormous sublattice asymmetry generates a quite large gap (~ 300 meV) in graphene, labelled as $E_{\sigma\uparrow}$ and $E_{\sigma\downarrow}$ for the spin-up and spin-down electrons respectively. Importantly, the strong π-d interaction between graphene and Ni not only opens up a large gap in graphene, but also lifts the degeneracy of the majority- and minority-spin bands around the K point of graphene, as shown in Fig. 3b. According to our calculation, the spin splitting of the graphene's conduction band is about 250 meV (similar for the valence band). The coexistence of the two effects (the gap opening and the spin splitting) induced by the Ni(111) substrate lead to the realization of the spin-polarized ferromagnetic semiconducting state with the gap of several tens meV in graphene, as schematically shown in Fig. 3c, which agrees well with our observations in the experiment. Since we observe the gap even at room-temperature, our results, therefore, indicate that the graphene monolayer on Ni(111) surface is the room-temperature ferromagnetic semiconductor. Very recently, the x-ray magnetic circular dichroism experiments also pointed out that the graphene/Ni(111) system can lead to magnetic moments in the carbon atoms of the graphene monolayer[44].

To further confirm the ferromagnetic semiconducting state of the graphene on Ni(111) surface, we carried out spin-polarized STM measurements, which provide us unprecedented opportunities to further identify the origin of the gap by directly detecting the majority- and minority-spin bands of graphene separately (Fig. 4)[45,46]. In our experiment, we used electrochemically etched Ni tips (Fig. S6 and Fig. S7) as the spin-polarized tip[47]. Due to the weak magnetocrystalline anisotropy of Ni[48], the magnetic polarization of the tip is always along the STM tip, i.e., perpendicular to the surface of the sample, due to the shape anisotropy of the STM tip. Before the STM measurements, a magnetic field of $B = 2.0$ T ($B = -2.0$ T) perpendicular to the surface of the sample was applied on and then removed gradually to obtain an up-polarized (down-polarized) STM tip. After that, a magnetic field $B = -0.02$ T, which is much smaller than the reversal field of the tip that usually ranges from about 200 to 300 mT in our experiments, is applied on the sample to make sure that the spins of the sample are down-polarized. As a consequence, the magnetizations of the STM tip and the sample are either parallel or antiparallel, as shown in Fig. 4a. Similar method can be adopted to make sure the spins of the sample up-polarized by using the magnetic field of $B = 0.02$ T, as shown in Figure 4b. Because of the spin conservation during the elastic electron tunnelling, we can detect the spin-up and spin-down bands of graphene separately, as schematically shown in Fig. 4. In the case that the magnetizations of the STM tip and the sample are parallel, the measured tunneling spectrum mainly reflects the local density-of-state (LDOS) of the spin-up electrons in the sample. In the other case that the magnetic polarizations of the tip and the sample are antiparallel, the recorded *dI/dV* spectrum is primarily contributed by the LDOS of the spin-down electrons in the sample. For simplicity, the above discussion are based on the assumption that the STM tip is 100% spin-polarized.

Fig. 4c shows four representative spin-resolved *dI/dV* spectra of the graphene monolayer on Ni (111) probed according to the above method (see Fig. S8-S11 for more experimental results). The spectra marked by green lines are measured when the spin polarizations of the tip and the sample are parallel, and the spectra marked by red lines are recorded when they are antiparallel. Very similar spectra have been obtained and verified by using several different Ni tips in our experiment. Additionally, we also rule out the DOS contributions of the underneath Ni(111) substrate by directly measuring the STS spectra on bare Ni (111) with spin-polarized Ni tip (see Fig. S12)[49]. The spin-polarized spectra recorded in the graphene monolayer on Ni (111) surface exhibit quite large gaps 200 ~ 300 meV, indicating that the strong interaction between graphene and Ni substrate really generates large gaps, $E_{\sigma\uparrow}$ and $E_{\sigma\downarrow}$, for both the spin-up and spin-down electrons in the graphene. Moreover, there is quite a large shift between the charge neutrality points (CNPs) of the spin-up and spin-down bands. The overlap between the $E_{\sigma\uparrow}$ and $E_{\sigma\downarrow}$ is much smaller than the gaps for either the spin-up or spin-down electrons. As a consequence, we can only detect a gap of several tens meV in graphene by using the non-magnetic STM tip, as shown in Fig. 4c. Our experimental result demonstrated explicitly that the joint effects of the gap opening and the spin splitting induced by the Ni substrate lead to the realization of the spin-polarized ferromagnetic semiconducting state in the graphene monolayer on Ni (111) surface.

In summary, we demonstrate, via spin-polarized STM measurements, that the strong interaction between graphene and Ni substrate generates a large gap in graphene and simultaneously leads to a relatively shift between the majority- and minority-spin bands. Consequently, the graphene sheet on Ni substrate exhibits a spin-polarized semiconductor band structure even at room temperature. Although the whole system is not a ferromagnetic semiconductor, our result indicates that it is facile to realize the room-temperature ferromagnetic state in a part of the artificial system. This may raise hopes of graphene-based novel technologies in the near future.

**Methods:**

**Density functional theory (DFT) Methods**

Density functional theory (DFT)[1] as implemented in the Vienna ab initio simulation package (VASP)[2] is performed for all the calculations. The local density approximation LDA[3] and the projector augmented wave (PAW)[4] are applied to describe the exchange-correlation energy and the core electrons, respectively. The plane-wave kinetic-energy cutoff is set at 400 eV. We construct a model of graphene-Ni (111) adsorption from a slab of six layers of Ni atoms with graphene adsorbed on one side. A vacuum layer of 12 A is used to keep the spurious interaction between neighboring slabs negligible. The lattice constant of graphene is chosen by its optimized LDA value, a=2.447 Å, adapting the lattice constant of the Ni (111) accordingly.

**STM/STS measurements**

The STM system was an ultrahigh vacuum scanning probe microscope (USM-1400S and USM-1500S ) from UNISOKU. All the STM and STS measurements were performed in the ultrahigh vacuum chamber (~$10^{-10}$ Torr) with constant-current scanning mode. The STM tips were obtained by chemical etching from a wire of Pt(80%) Ir(20%) alloys. Lateral dimensions observed in the STM images were calibrated using a standard graphene lattice and a Si (111)-(7 ×7) lattice and Ag (111) surface. The dI/dV measurements were taken with a standard lock-in technique by turning off the feedback circuit and using a 793-Hz 5 mV a.c. modulation of the sample voltage. All the STS measurements were not recorded until the right standard tunnelling spectra of graphite was obtained.

**LPCVD method to grow graphene on Ni foils.**

A traditional ambient pressure chemical vapor deposition (APCVD) method was adopted to grow controllable layers of graphene on Ni(111) single crystal. The Ni(111) single crystal (5 mm × 5 mm width, 1 mm thickness) was first heated from room temperature to 900℃ in 40 min under an argon (Ar) flow of 100 SCCM (SCCM stands for Standard Cubic Centimeters per Minute) and hydrogen ($H_2$) flow of 100 SCCM, and keep this temperature and flow ratio for 20 min. Next methane ($CH_4$) gas was introduced with a flow ratio of 5 SCCM, the growth time is ~15 min, and then cooled down to room temperature (see Figure S1). Then samples are transferred into the ultrahigh vacuum condition for further characterizations.

**Acknowledgments**

This work was supported by the National Natural Science Foundation of China (Grant Nos. 11674029, 11422430, 11374035), the National Basic Research Program of China (Grants Nos. 2014CB920903, 2013CBA01603), the program for New Century Excellent Talents in University of the Ministry of Education of China (Grant No. NCET-13-0054), the China Postdoctoral Science Foundation (No. 212400207). L.H. also acknowledges support from the National Program for Support of Top-notch Young Professionals and support from "the Fundamental Research Funds for the Central Universities".

**Author contributions**

Y.Z., D.L.M. and K.K.B synthesized the samples and performed the experiments. L.H. and Y.Z. analyzed the data. X.L.S. and W.H.D. performed the theoretical calculations. L.H. conceived and provided advice on the experiment, analysis, and theoretical calculation. L.H. and Y.Z. wrote the paper. All authors participated in the data discussion. All authors participated in the data discussion.

**Additional information**

Supplementary information is available in the online version of the paper. Reprints and permissions information is available online at www.nature.com/reprints. Correspondence and requests for materials should be addressed to L.H.

**Competing financial interests**

The authors declare no competing financial interests.

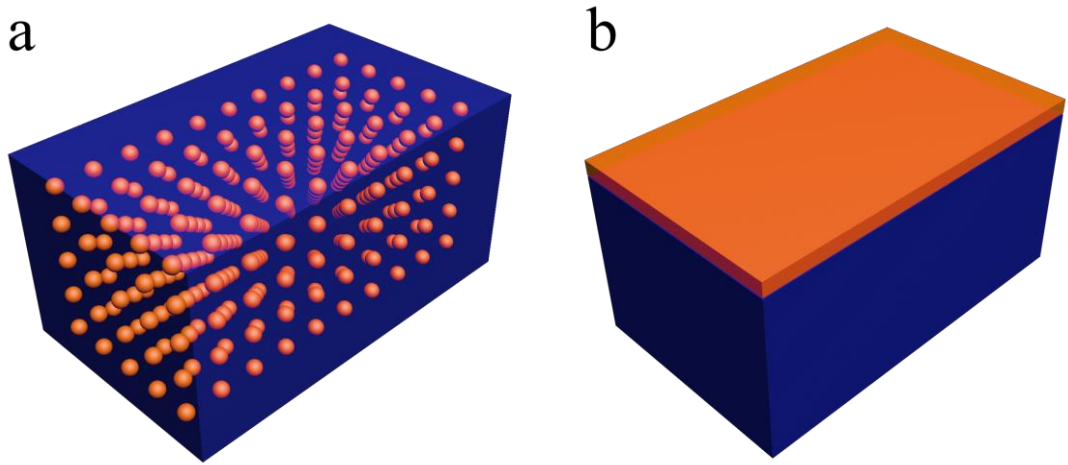

**Fig. 1. Two different methods to realize room-temperature ferromagnetic semiconducting state. a.** Realizing room-temperature ferromagnetic semiconductor state in the whole ultimate material: it could be achieved either by synthesizing gapped materials with high $T_C$ ferromagnetism or by introducing magnetic ions into nonmagnetic semiconductors. **b.** A new recipe was proposed to achieve room-temperature ferromagnetic semiconducting state by only driving a small part, for example, only the surface, of an artificial system into the room-temperature ferromagnetic semiconductor state.

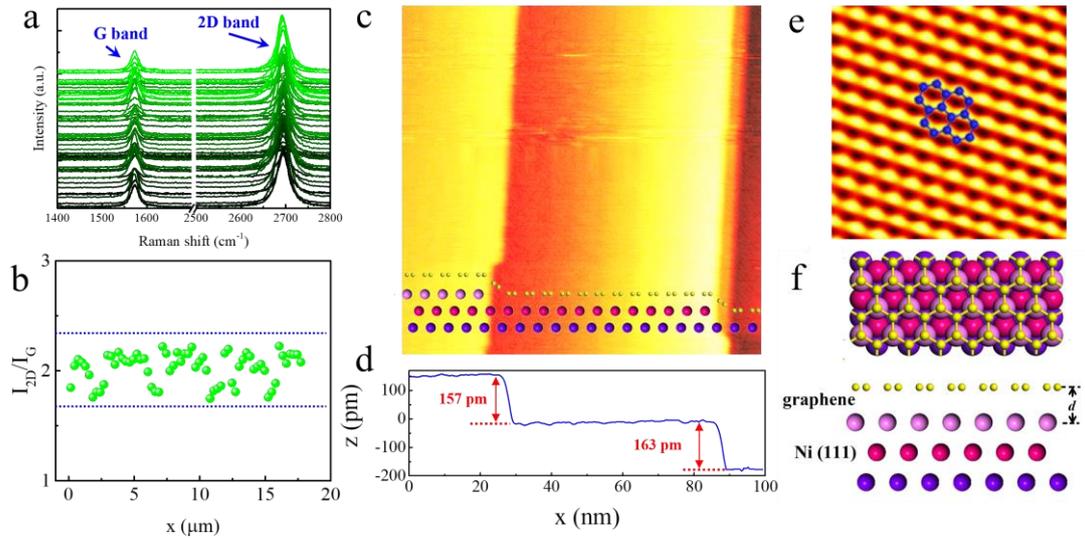

**Fig. 2. Graphene monolayer on Ni(111) single crystal. a.** The Raman spectra of graphene transferred to a 300 nm SiO$_2$/Si substrate measured at different positions. The peaks at ~2700 cm$^{-1}$ and ~1580 cm$^{-1}$ of the Raman spectra are the 2D band and G band of graphene, respectively. **b.** The ratios of the intensities of 2D band and G band vary with the recorded positions. The average value of $I_{2D}/I_G \sim 2$ indicates that the synthesized sample is graphene monolayer. **c.** $100\ nm \times 100\ nm$ STM topographic images of graphene monolayer across typical monatomic Ni(111) steps ($V_b = 0.5\ V$, $I = 0.2\ nA$). **d.** The height profile along the monatomic Ni(111) steps in c. **e.** Zoom-in atomic resolution STM image of graphene, showing the triangular lattice of the monolayer graphene on the Ni(111) surface ($V_b = 0.5\ V$, $I = 0.2\ nA$). **f.** Schematic image of a graphene monolayer on Ni(111) surface. The carbon atoms of the *A* sublattice in graphene are on top of the Ni atoms of the topmost atomic layer, while atoms of the *B* sublattice are located above the hollow sites of the topmost atomic layer of Ni(111) surface.

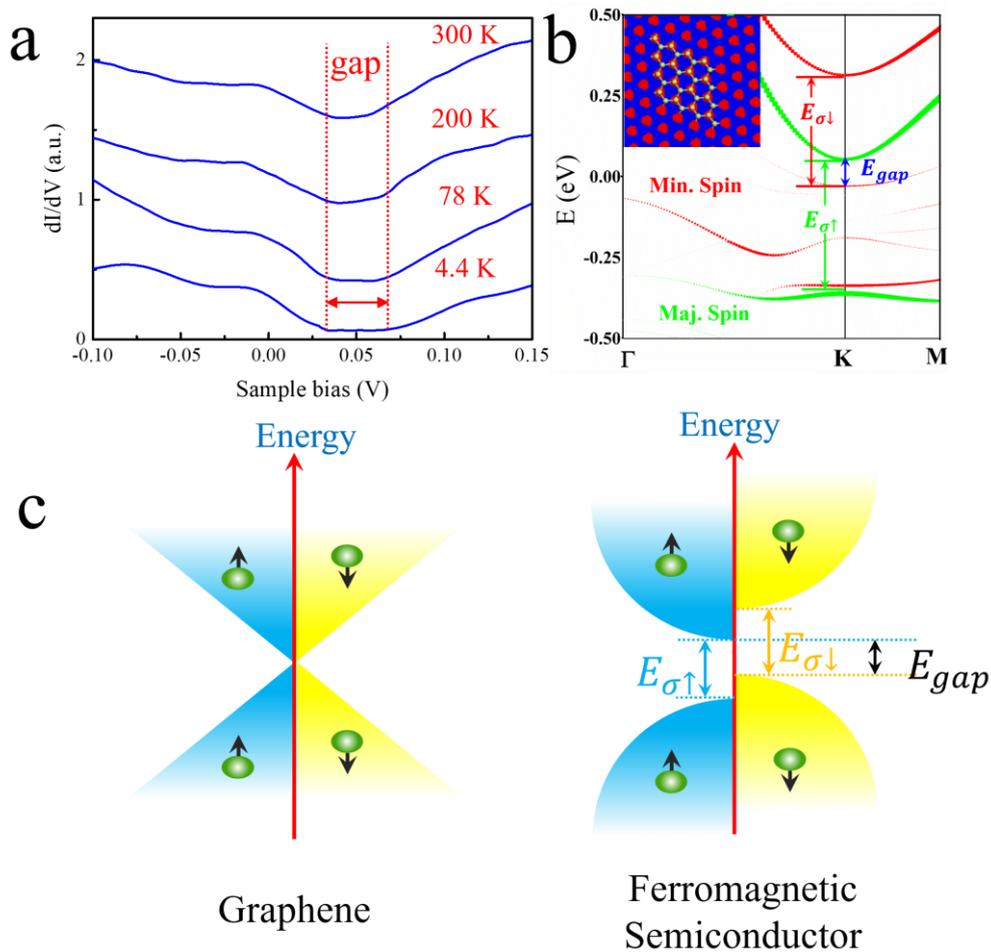

**Fig. 3. Microscopic properties of the graphene monolayer on Ni(111) surface. a.** Representative STS spectra of the graphene on Ni(111) surface recorded at 4.4 K, 78 K, 200 K and 300 K, respectively. The STS spectra are offset for clarity. **b.** Band structures of graphene monolayer adsorbed upon Ni(111) surface. The Fermi level is at zero energy. The labels Maj. Spin and Min. Spin indicate the majority-spin bands (green lines) and minority-spin bands (red lines) of the graphene monolayer on Ni(111), respectively. Inset: DFT-simulated STM image of the graphene monolayer on Ni(111) surface. **c.** (Left) Low-energy electronic band structure of graphene monolayer. The electronic structures for the spin-up and spin-down states are degenerate. (Right) Schematic band structure of the ferromagnetic semiconductor. In this work, strong interaction between graphene and a magnetic metal (Ni) leads to a band gap and spin splitting simultaneously in graphene, which makes the graphene monolayer a ferromagnetic semiconductor.

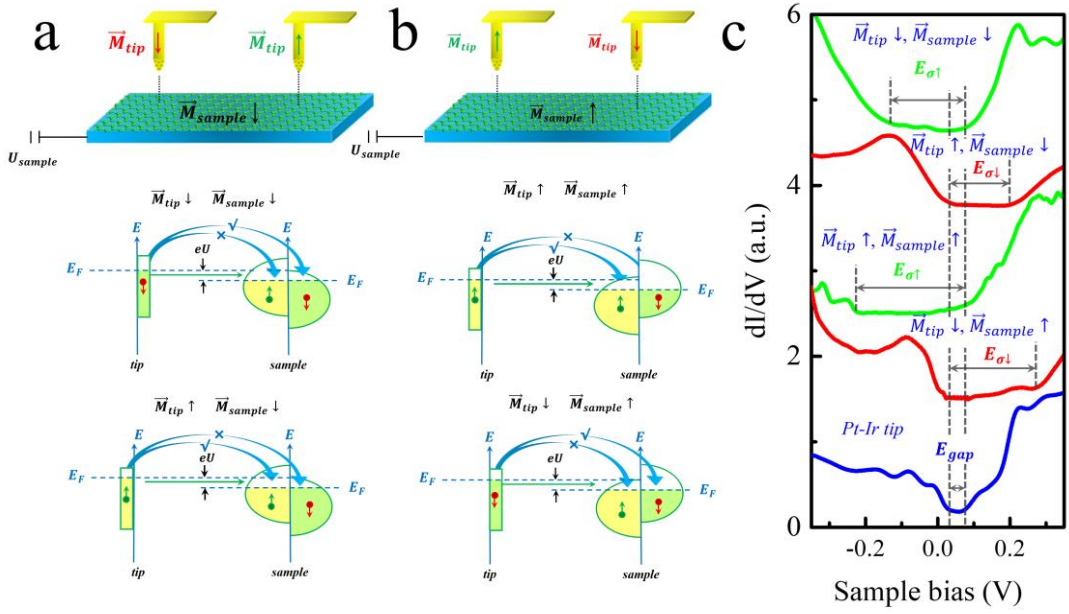

**Fig. 4. Spin-polarized dI/dV spectra of the graphene monolayer on Ni (111) surface.**
(**a, b**) Top panel: schematic experimental set-up of spin-polarized STM. Spins of the sample aligned perpendicular to the sample surface can be detected by the Ni tip along the tip axis. Bottom panels: principle of spin-polarized tunnelling between magnetic electrodes that exhibit a parallel ($\vec{M}_{tip} \downarrow$, $\vec{M}_{sample} \downarrow$ or $\vec{M}_{tip} \uparrow$, $\vec{M}_{sample} \uparrow$) and an antiparallel ($\vec{M}_{tip} \uparrow$, $\vec{M}_{sample} \downarrow$ or $\vec{M}_{tip} \downarrow$, $\vec{M}_{sample} \uparrow$) magnetization. The spin is conserved during the elastic electron tunnelling. (**c**) Spin-polarized *dI/dV* spectra of the graphene monolayer on Ni surface. The red curve is measured when the tip polarization $\vec{M}_{tip}$ and the magnetization of the sample $\vec{M}_{sample}$ are antiparallel, which reflects the LDOS of the spin-down electrons. The green curve is measured when the $\vec{M}_{tip}$ and $\vec{M}_{sample}$ are parallel, which reflects the LDOS of the spin-up electrons. The blue dots are the dI/dV spectrum of graphene on Ni(111) surface measured via a nonmagnetic Pt/Ir tip.